\documentclass{article}

\usepackage{arxiv}

\usepackage[utf8]{inputenc} 
\usepackage[T1]{fontenc}    
\usepackage{hyperref}       
\usepackage{url}            
\usepackage{booktabs}       
\usepackage{amsfonts}       
\usepackage{nicefrac}       
\usepackage{microtype}      
\usepackage{lipsum}
\usepackage{graphicx}
\graphicspath{ {./images/} }

\usepackage{graphicx}
\usepackage[caption = false]{subfig}
\usepackage[round]{natbib}
\usepackage{amsmath}  
\usepackage{graphicx} 
\usepackage{mathtools}
\usepackage{algorithm}
\usepackage[noend]{algpseudocode}
\usepackage{float}
\usepackage[toc,title,page]{appendix}

\newcommand{\Real}{\mathbb R}

\renewcommand{\P}{\mathbb{P}}

\newcommand{\eps}{\varepsilon}

\newcommand{\F}{\mathcal{F}}

\newcommand{\one}[1]{\mathbf{1}_{\{#1\}}}

\newcommand{\D}{\mathrm{d}}

\newcommand{\ben}{\begin{enumerate}}
\newcommand{\een}{\end{enumerate}}

\renewcommand{\vec}[1]{\mathbf{#1}}

\newcommand*\diff{\mathop{}\!\mathrm{d}}

\DeclareMathOperator{\esssupp}{ess\,supp}

\title{Sequential Importance Sampling With Corrections For Partially Observed States}

\author{
 Valentina Di Marco  \\
  School of Mathematical Science\\
  Monash University, Clayton Campus\\
  VIC 3800, Australia \\
  \texttt{valentina.dimarco@monash.edu} \\
   \And
 Jonathan Keith \\
  School of Mathematical Science\\
  Monash University, Clayton Campus\\
  VIC 3800, Australia \\
  \texttt{jonathan.keith@monash.edu} \\
}

\begin{document}
\maketitle
\begin{abstract}
We consider an evolving system for which a sequence of observations is being made, with each observation revealing additional information about current and past states of the system. We suppose each observation is made without error, but does not fully determine the state of the system at the time it is made. 

Our motivating example is drawn from invasive species biology, where it is common to know the precise location of invasive organisms that have been detected by a surveillance program, but at any time during the program there are invaders that have not been detected.

We propose a sequential importance sampling strategy to infer the state of the invasion under a Bayesian model of such a system. The strategy involves simulating multiple alternative states consistent with current knowledge of the system, as revealed by the observations. However, a difficult problem that arises is that observations made at a later time are invariably incompatible with previously simulated states. To solve this problem, we propose a two-step iterative process in which states of the system are alternately simulated in accordance with past observations, then corrected in light of new observations. We identify criteria under which such corrections can be made while maintaining appropriate importance weights.
\end{abstract}

\keywords{Sequential Importance Sampling \and Filtering \and Bayesian estimation \and Partially observed spaces \and Missing data}

\section{Introduction}
\label{sec:1}

This paper considers the problem of imputing missing data in the presence of incomplete observations made sequentially in time. We thus envisage data that consists of a series of correlated observations made sequentially, each of which is correct but only partially reveals the true state of the system. The main difficulty that arises in this context is that data missing at one time point can be revealed at a later time point, so that imputed missing values must later be corrected in light of new information.

Our original motivation for considering this problem was to facilitate analysis of invasive species, where in an ongoing surveillance program the locations of invaders are regularly being detected. For detected invaders, the location can be precisely determined, but at any given time there is an unknown number of undetected organisms, each with an unknown location. To infer the current extent of the invasion, we aim to impute plausible locations of undetected individuals, but these imputations are only informed guesses, and will require constant correction as new observations come to light. In addition, new nests are constantly being produced: the unseen state of the system is thus constantly evolving. Knowing the location of at least some of the invaders at some time $t$, we can simulate the evolution of this system. However, again the imputed locations of simulated nests will require constant correction as more information about the true state of the system becomes available.

Here we propose a new approach to problems of this type. Although our approach is ultimately aimed at inference for invasive species, we will illustrate the method for less complex evolving systems in which correct but partial observations are being made in real time.

In the first chapter of this paper we introduce our Bayesian approach that uses a new sequential importance sampling (SIS) strategy. As is typical of SIS methods, we generate a population of particles, each representing a plausible sequence of system states, and we evolve each particle at each time step according to a model of system dynamics. Again in a manner typical of SIS methods, new observations arrive in real time, and we use these observations to adjust the weights assigned to particles. However, a crucial new element in our method is that we allow missing values imputed at earlier time steps to be corrected so that they are consistent with the new observations. We test our method with two simple models in chapter 2 and 3. Firstly, we apply the method to an AR(1) model where we simulate a set of observations then reconstruct the system. These results are subsequently compared with the analytical solutions obtained directly applying the AR(1) model definition. In order to test the applicability of our method to a simple invasive species problem, we therefore introduce a model for a partially observed river invasion and impute the missing data.

Our method addresses a problem that is common in practice: new observations made in real time can render previously imputed missing values implausible, and may even conflict with what has been simulated. For example, in the invasive species context described above, undetected individuals imputed to geographic areas that do not contain any invaders become increasingly implausible as time passes without any detection being made in those regions. In standard SIS approaches, such particles receive low weights and are eventually eliminated by resampling, but a common problem is that all particles can become implausible, if not inconsistent, with observations.

Problems of this kind arise in many contexts other than invasive species detection. We can envisage the approach being used to study the evolution of a species' geographic range, for both invasive and non-invasive species. The algorithm could also be applied to a variety of other missing data problems where new incomplete data is continuously acquired, such as in the prediction of earthquake aftershocks, (\cite{Seif}) or bushfire modeling (\cite{Beer}). Another potential application is in the study of the spread of infectious diseases, where observations are made in the form of diagnosed cases, but there is missing data in the form of undiagnosed cases (\cite{O'Neill}). 

Missing data problems are ubiquitous in ecological and evolutionary data sets as in many other branches of science. Two common methods used to deal with missing data are to delete rows of a data matrix (corresponding to individuals or cases) that contain missing data, or to use the mean to fill in missing values. However, these methods result in biased estimation of parameters and uncertainty, and reduction in statistical power. Better missing data procedures such as data augmentation (\cite{Tanner}) and multiple imputation (\cite{RubinMI}) are available  (\cite{Nakagawa}).
For an up-to-date treatment of missing data in statistics see \cite{Little}.
However, techniques like multiple imputations and data augmentation generally do not make use of past observations and the state transition equation of the system when estimating the probability of the hidden state in light of the observations. This can result in poor performance when the problem can be well modelled, for example, by a Markov structure (\cite{Zhang}).

An example of an alternative methodology for dealing with missing data in the context of a Markov structure is the Multiple Imputation Particle Filter (MIPF) developed by \cite{Zhang}. This method, applied to signal processing, uses randomly drawn values (imputations) to provide a replacement for the missing data and then uses a particle filter to estimate non-linear state with the data.

Particle filters are sometimes used to estimate a hidden state when partial, noisy observations are made. However, the performance of particle filtering algorithms can severely degrade in the presence of missing data (\cite{Zhang}). For a comprehensive discussion of the applications of particle filters, also called Sequential Monte Carlo (SMC) methods, see \cite{Cappe} and \cite{Doucet}.

Sequential importance sampling (SIS) is a form of Particle filter firstly introduced by \cite{Kong} that has been used in Bayesian missing data problems in the form of sequential imputation, in situations where the posterior must be constantly updated with the arrival of new data.

Here we propose a two-step iterative procedure for dynamically updating a collection of weighted particles used to represent a posterior distribution over the possible trajectories of a partially observed system. In the first step, we evolve particles in accordance with a model of the system, in a manner common to particle-based sequential Monte Carlo methods. In the second step, we correct the particles in light of newly acquired data. The method treats an uncorrected particle (generated in the first step) as a form of augmented variable. Thus the method constitutes a novel use of augmentation in sequential importance sampling. 

As mentioned above, the observations in the problems we consider are partial, but exact. In general, in situations where the data are highly informative standard sequential Monte Carlo methods can perform poorly. 

\cite{Del Moral} proposed a Sequential Monte Carlo method for sampling the posterior distribution of state-space models under highly informative observation regimes. In their method they introduced a schedule of intermediate weighting and resampling times between observation times, which guides particles towards the final state.

\cite{Finke} developed a Particle Monte Carlo Markov Chain algorithm to estimate the demographic parameters of a population and then incorporated this algorithm into a sequential Monte Carlo sampler in order to perform model comparison motivated by the fact that a simple importance sampling performs poorly if there is a strong mismatch between the prior and the posterior, which is common when the data is highly informative.

Our new method gives similar results to the gold standard and is able to handle well the reconstruction of a river invasion. The cancellations we get during the calculation of the weights are a pleasing feature of the method, but we have yet to test the full potential and speed of the algorithm when the data to analyse is substantial.

\section{Sequential Importance Sampling with corrections}
\label{sec:2}

In this chapter we present the method in a general context for an $M$-dimensional system, evolving in discrete time. The states of this system at times $t = 1, \dots, T$ are represented by random vectors $\vec{x}^t = (x^t_1, \dots, x^t_M) \in \Real^M$ defined on a probability space $(\Omega, \F, \P)$. The trajectory of the system up to time $t$ we represent by a random matrix $\vec{X}^t = (\vec{x}^1, \ldots, \vec{x}^t)$. We also represent our state of knowledge regarding the trajectory of the system up to time $t$ by a random binary matrix $\vec{b}^t = (b^t_{im}) \in 2^{t \times M}$, where $b^t_{im} = 1$ if the value of $x^i_m$ is known by time $t$, and $b^t_{im} = 0$ if the value of $x^i_m$ is still unknown at time $t$. Note that once a past system coordinate $x^i_m$ is known, it cannot become unknown, so $b^t_{im} = 1$ implies $b^{t^\prime}_{im} = 1$ at all later times $t^{\prime} > t$, and similarly $b^t_{im} = 0$ implies $b^{t^{\prime}}_{im} = 0$ at all earlier times $t^{\prime} < t$. It will also be convenient to define $\vec{B}^t = (\vec{b}^1,\ldots,\vec{b}^t)$.

We define an {\em observation matrix} $\vec{z}^t = (z^t_{im})$, where $z^t_{im} = x^i_m$ if $b^t_{im} = 1$, and $z^t_{im} = -$ if $b^t_{im} = 0$. Thus the symbol `$-$' is used to represent an unknown system coordinate. It will also be convenient to define $\vec{Z}^t = (\vec{z}^1,\ldots,\vec{z}^t)$. 

We assume observations are made without error, so that $\vec{Z}^t$ is fully determined by $\vec{X}^t$ and $\vec{B}^t$. We therefore define a function $\sigma_t: \Real^{t \times M} \times \left( \prod_{i=1}^t 2^{i \times M} \right) \rightarrow \prod_{i=1}^t (\Real \cup \{ - \})^{i \times M}$ such that $\vec{Z}^t = \sigma_t(\vec{X}^t, \vec{B}^t)$. (In what follows, the subscript on $\sigma$ is omitted, as it is implied by the superscripts on the arguments.) Note that $\vec{Z}^t$ fully determines $\vec{B}^t$ and those elements $x_m^i$ of $\vec{X}^t$ for which $b_{im}^t = 1$, but leaves the remaining elements of $\vec{X}^t$ undetermined.

For each past system coordinate $x^i_m$ that has not been observed by time $t$ (so that $b^t_{im} = 0$) there may nevertheless be some information that can be derived from the observations and a model of the system. We take a Bayesian approach to quantify this information. At time $t=1$, our knowledge of the system is represented by a prior distribution with density $q(\vec{X}^1, \vec{B}^1)$ on $\Omega_1 = \Real^M \times 2^M$ and the posterior distribution after observing $\vec{Z}^1$ has a density on $\Omega_1' = \sigma^{-1}(\vec{Z}^1)$ given by:
\begin{equation*}
    p(\vec{X}^1, \vec{B}^1 |\vec{Z}^1) =
        \frac{q(\vec{X}^1, \vec{B}^1)}
        {r(\vec{z}^1)} 
\end{equation*}
where
\begin{align*}
    r(\vec{z}^1)  &= \int_{\Omega_1'} q(\vec{X}^1, \vec{B}^1) \prod_{ \{m \in \{1, \ldots, M \} : z^1_{1m} = - \} } \D x^1_m.
\end{align*}
Note that the preceding equation integrates over the elements of $\vec{X}^1$ that are not determined by $\vec{Z}^1$. Thus $r(\vec{z}^1)$ is the integral of $q(\vec{X}^1, \vec{B}^1)$ over $\Omega_1'$. 

Our knowledge of the trajectory of the system at time $t \geq 2$, before we acquire the next observation matrix $\vec{z}^t$, is represented by a prior distribution with density $q(\vec{X}^t, \vec{B}^t | \vec{Z}^{t-1})$ on the subspace $\Omega_t = \sigma^{-1}(\vec{Z}^{t-1}) \times \Real^M \times 2^{t \times M}$. After observing $\vec{z}^t$, certain pairs $(\vec{X}^t, \vec{B}^t)$ with non-zero prior density will be incompatible with the new observations, and the posterior density must therefore restrict $(\vec{X}^t, \vec{B}^t)$ to $\Omega_t' = \sigma^{-1}(\vec{Z}^t)$. The posterior density over $\Omega_t'$ must therefore be
\begin{equation*}
    p(\vec{X}^t, \vec{B}^t |\vec{Z}^t) =
        \frac{q(\vec{X}^t, \vec{B}^t | \vec{Z}^{t-1})}
        {r(\vec{z}^t | \vec{Z}^{t-1})} 
\end{equation*}
where
\begin{align*}
    r(\vec{z}^t | \vec{Z}^{t-1})  &= \int_{\Omega_t'} q(\vec{X}^t, \vec{B}^t | \vec{Z}^{t-1}) \prod_{ \{ i \leq t, m \in \{1, \ldots, M \} : z^t_{im} = - \} } \D x^i_m.
\end{align*}
Note that the preceding equation integrates over the elements of $\vec{X}^t$ that are not determined by $\vec{Z}^t$. Thus $r(\vec{z}^t | \vec{Z}^{t-1})$ is the integral of $q(\vec{X}^t, \vec{B}^t | \vec{Z}^{t-1})$ over $\Omega_t'$. 

In this paper we focus on Markovian systems, that is, systems in which each $\vec{x}^t$ is conditionally independent of $\vec{X}^{t-2} = (\vec{x}^1, \ldots,\vec{x}^{t-2})$, given $\vec{x}^{t-1}$. Let $f(\vec{x}^t | \vec{x}^{t-1})$ be the transitional density distribution characterising the system. We also suppose that our current state of knowledge regarding the trajectory of the system up to and including the current time depends on both the trajectory of the system and our knowledge of it at the previous time, as follows:
\begin{align*}
    &\vec{x}^t | \vec{x}^{t-1} \sim f_t(\vec{x}^t | \vec{x}^{t-1})\\
    &\vec{b}^t | \vec{x}^t, \vec{b}^{t-1} \sim g_t(\vec{b}^t | \vec{x}^t, \vec{b}^{t-1})
\end{align*}
for $t \geq 2$, with densities of the initial states $f_1(\vec{x}^1)$ and $g_1(\vec{b}^1 | \vec{x}^1)$. (In what follows, the subscripts on $f$ and $g$ are omitted, as they are implied by the superscripts on the arguments.)
This general framework allows for the possibility that $\vec{b}^t$ is independent of $\vec{x}^{t}$ and depends only on $\vec{b}^{t-1}$. It also allows for the alternative possibilities that observations are made in response to what is known about the trajectory of the system and/or that the probability of making an observation depends on the trajectory of the system. Thus the joint prior distribution for $(\vec{X}^t,\vec{B}^t)$ when $t=1$ is given by 
\[
q(\vec{X}^1,\vec{B}^1) = f(\vec{x}^1)g(\vec{b}^1 | \vec{x}^1)
\]
and when $t \geq 2$: 
\begin{eqnarray*}
    q(\vec{X}^t, \vec{B}^t | \vec{Z}^{t-1}) &=& f(\vec{x}^t | \vec{x}^{t-1}) g(\vec{b}^t | \vec{x}^t, \vec{b}^{t-1}) p(\vec{X}^{t-1}, \vec{B}^{t-1} | \vec{Z}^{t-1}) \\
& = & \frac{f(\vec{x}^1)g(\vec{b}^1 | \vec{x}^1)}{r(\vec{z}^1)} \prod_{i=2}^{t-1} \frac{f(\vec{x}^i | \vec{x}^{i-1}) g(\vec{b}^i | \vec{x}^i, \vec{b}^{i-1})}{r(\vec{z}^i | \vec{Z}^{i-1})} \\
& & \times f(\vec{x}^t | \vec{x}^{t-1}) g(\vec{b}^t | \vec{x}^t, \vec{b}^{t-1})
\end{eqnarray*}
on the space $\Omega_t$, and the joint posterior distribution is given by:
\begin{eqnarray*}
    p(\vec{X}^t, \vec{B}^t | \vec{Z}^t) 
& = & \frac{f(\vec{x}^1)g(\vec{b}^1 | \vec{x}^1)}{r(\vec{z}^1)} \prod_{i=2}^t \frac{f(\vec{x}^i | \vec{x}^{i-1}) g(\vec{b}^i | \vec{x}^i , \vec{b}^{i-1})}{r(\vec{z}^i | \vec{Z}^{i-1})} 
\end{eqnarray*}
on $\Omega_t'$. It is also helpful to define
\begin{eqnarray*}
r(\vec{Z}^t) & := & r(\vec{z}^1) \prod_{i=2}^{t} r(\vec{z}^i | \vec{Z}^{i-1}) \\
& = & \int_{\Omega_t'}  h(\vec{X}^t,\vec{B}^t) \prod_{ \{ i \leq t, m \in \{1, \ldots, M \} : z^t_{im} = - \} } \D x^i_m.
\end{eqnarray*}
where
\begin{eqnarray*}
h(\vec{X}^t,\vec{B}^t) & := & f(\vec{x}^1)g(\vec{b}^1 | \vec{x}^1) \prod_{i=2}^{t} f(\vec{x}^i | \vec{x}^{i-1}) g(\vec{b}^i | \vec{x}^i , \vec{b}^{i-1}).
\end{eqnarray*}
We want to approximate the distribution $p(\vec{X}^{t}, \vec{B}^{t} | \vec{Z}^{t})$ by iterative sampling, and thus obtain Monte Carlo estimates of expectations of the form:
\begin{align*}
    &E_{p} \Big[l(\vec{X}^{t}, \vec{B}^{t} ) \Big| \vec{Z}^{t} \Big] = \\
    &\int_{\Omega_t'} l(\vec{X}^{t}, \vec{B}^{t})\; p(\vec{X}^{t}, \vec{B}^{t} | \vec{Z}^{t})\; \prod_{ \{ i \leq t, m \in \{1, \ldots, M \} : z^t_{im} = - \} } \D x^i_m
\end{align*}
for functions $l: \Omega_t' \rightarrow \Real$. 

We take a sequential importance sampling approach, at each iteration using a collection of weighted particles to approximate $p(\vec{X}^{t-1}, \vec{B}^{t-1} | \vec{Z}^{t-1})$, in the sense that these particles can be used to construct weighted Monte Carlo estimates for integrals of the above form. For simplicity, we suppose particles are resampled at the end of each iteration, thus resetting the particle weights to be equal. However, the methods we describe below can be modified in a straightforward way if resampling is not performed, in which case the weights from successive iterations multiply, as is usual in sequential importance sampling.

At each time point, we evolve the particles under the model to create a new set of particles representing the prior for iteration $t$, namely $q(\vec{X}^{t}, \vec{B}^{t} | \vec{Z}^{t-1})$. We then modify these particles to be consistent with the observations $\vec{Z}^{t}$, and adjust the weights to ensure the resulting weighted particles provide consistent importance sampling estimates of an expectation $E_{p}[l(\vec{X}^{t}, \vec{B}^t) | \vec{Z}^t]$.

Consider a particle $(\vec{X}^{t-1}_j,\vec{B}^{t-1}_j)$ constructed at time point $t$, for $j = 1, \ldots, n$, where $n$ is a fixed number of particles. We generate $\vec{x}^t_j$ for this particle by sampling from $f(\vec{x}^t | \vec{x}^{t-1}_j)$. Similarly, we generate $\vec{b}^{t}_j$ for this particle by sampling from $g(\vec{b}^{t} | \vec{x}^t_j, \vec{b}^{t-1}_j)$. However, the new matrix of observations $\vec{z}^{t}$ is typically inconsistent with a particle $(\vec{X}^{t}_j, \vec{B}^{t}_j)$ thus constructed, in two ways. First, the coordinates at which $\vec{b}^{t}_j$ contains a 0 may not correspond to the coordinates at which $\vec{z}^{t}$ contains a `$-$' and coordinates at which $\vec{b}^{t}_j$ contains a 1 may not correspond to an observation in $\vec{z}^{t}$. Second, the observed values in $\vec{z}^{t}$ may differ from the corresponding coordinates of $\vec{X}^{t}_j$. We must therefore correct $\vec{X}^{t}_j$ and $\vec{b}^{t}_j$ in light of the new observations $\vec{z}^{t}$. The simplest way to do this is to first replace $\vec{b}^{t}_j$ with the unique binary matrix $(\vec{b}_j')^{t}$ that is consistent with $\vec{z}^{t}$, and then replace the coordinates of $\vec{X}^{t}_j$ with the corresponding coordinates of $\vec{z}^{t}$ wherever $(\vec{b}_j')^{t}$ has a `1', thus generating a corrected term $\vec{(X')}^{t}_j$. The corrections thus made at time point $t$ will be carried forward into the particles used at all future times. Here we use deterministic corrections, in which the simulated pair $\vec{y}^{t} = (\vec{X}^{t}, \vec{B}^{t}) \in \Omega_t$ can be corrected in only one way to produce a new element $\vec{(y')}^{t} = (\vec{(X')}^t, \vec{(B')}^t) \in \Omega_t'$ (note we have repressed the particle subscript $j$). That is, 
\[
\vec{(y')}^{t} = \rho(\vec{y}^{t},\vec{Z}^{t})
\]
for some function $\rho$. The technique we propose can be generalised for non-deterministic corrections in a natural way, though we shall not do so here.

At each iteration, in order to simulate via a two-step process in which we generate a sample and then correct in light of data, we propose to use an auxiliary variable in the following manner. The auxiliary variable will be the yet to be corrected sample $\vec{y}^{t}$, which is an element of the space $\Omega_t$. We will use a projection map $\pi$ to relate the augmented space $\Omega^*_t = \Omega_t \times \Omega_t'$ containing elements of the form $(\vec{y}^{t}, \vec{(y')}^{t})$ to the corrected state space $\Omega_t'$ containing elements of the form $\vec{(y')}^{t}$. Our strategy is to define a probability density $p^*$ on $\Omega^*_t$, such that the marginal density of $p^*$ on $\Omega_t'$ is $p$. Similarly, we define a density $q^*$ on $\Omega^*_t$, such that the marginal density of $q^*$ on $\Omega_t$ is $q$. We then use a sequential importance sampling approach to re-weight a sample of particles used for Monte Carlo estimation with respect to $q^*$, so that they can be used for Monte Carlo estimation with respect to $p^*$.

A key identity underlying this strategy is the following.
\[
E_{p}[l(\vec{(y')}^{t})]  =  E_{p^*}[l(\pi (\vec{y}^{t}, \vec{(y')}^{t})]
\]
In this equation, the expectation on the left is over $\Omega_t'$, whereas the expectation on the right is over $\Omega_t^*$. We sketch a proof of this identity in the appendix.

Since we are here interested in a deterministic correction, we define both $p^*$ and $q^*$ on a subspace of $\Omega_t^*$ consisting of points of the form $(\vec{y}^{t}, \rho(\vec{y}^{t}, \vec{Z}^{t}))$. (The densities $p^*$ and $q^*$ are defined with respect to a reference measure on this subspace, constructed with the aid of Rohlin's disintegration theorem (\cite{Rohlin}), though we shall not provide the measure theoretic details here.) Since $\vec{y}^{t}$ determines $\vec{(y')}^{t}$, but the reverse is not necessarily true, the new densities are of the form:
\begin{align*}
q^*(\vec{y}^{t},\vec{(y')}^{t}) &= q(\vec{y}^{t} | \vec{Z}^{t-1}), \mbox{ and } \\
p^*(\vec{y}^{t},\vec{(y')}^{t}) &= u(\vec{y}^{t} | \vec{(y')}^{t}) p(\vec{(y')}^{t} | \vec{Z}^{t})
\end{align*}
where $\vec{(y')}^{t} = \rho(\vec{y}^{t}, \vec{Z}^{t})$. (The conditioning on $\vec{Z}^{t-1}$ and $\vec{Z}^{t}$ has been suppressed on the left hand side of these definitions.) Here $u$ is the density of some distribution over the set 
\[
F(\vec{(y')}^{t}) = \{ \vec{y} \in \Omega^t  :  \rho(\vec{y}, \vec{Z}^{t}) = \vec{(y')}^{t} \}.
\]
Note the right hand side of the above definition depends on $\vec{Z}^{t}$, but this is not shown as an argument on the left hand side because $\vec{Z}^{t}$ is uniquely determined by $(\vec{y}')^{t}$.

The set $F(\vec{(y')}^{t})$ contains all elements of $\Omega^t$ that can be corrected to $\vec{(y')}^{t}$ by the above procedure (including $\vec{y}^{t}$ and $\vec{(y')}^{t}$ itself). It can be characterised as the set of elements of the form $(\vec{X}^{t},\vec{B}^{t}) \in \Omega^t$ such that:
\begin{enumerate}
\item $\vec{b}^{j} = (\vec{b}')^{j}$ for all $j < t$,
\item $b_{im}^{t} = (b_{im}')^t$ whenever $(b_{im}')^t = (b_{im}')^{t-1}$, otherwise $b_{im}^{t} \in \{ 0, 1 \}$, and
\item $x_{m}^{i} = (x_{m}')^{i}$ whenever $(b_{im}')^t = (b_{im}')^{t-1}$, otherwise $x_{m}^{i} \in \Real$.
\end{enumerate}
Thus $F(\vec{(y')}^{t})$ is isomorphic to $\Real^k \times 2^M$, where $k$ is the number of coordinates of $\vec{B}^{t}$ at which $(b_{im}')^t \neq (b_{im}')^{t-1}$, that is, the number of newly observed coordinates.

The density $u$ is somewhat flexible, but must ensure $\esssupp(p^*) \subseteq \esssupp(q^*)$, which is a crucial requirement for valid importance sampling (\cite{Geweke}). Here we consider two alternatives for $u$. The first, $u_1$, is the uniform density on a bounded subset of $F(\vec{(y')}^{t})$.  The subset must be bounded to ensure $u$ is integrable (with respect to the appropriate reference measure). If the bounded set is a hyper-rectangle, with each undetermined coordinate of $\vec{y}^{t}$ bounded independently of the others, then the normalising constant of $u_1$ depends on the number of newly observed coordinates $k$, but is otherwise independent of $\vec{(y')}^{t}$. If $\vec{y}^t$ is outside this hyper-rectangle, then $u_1(\vec{y}^t) = 0$, and as we shall see later in this section, this results in a zero-weighted particle in the context of importance sampling. In practice, the bounds can be set large enough that zero-weighted particles occur only infrequently, and cause no problems. The second alternative, $u_2$, is obtained by restricting $q(\vec{y}^{t} | \vec{Z}^{t-1})$ to the set $F(\vec{(y')}^{t})$. In this case, the normalising constant, obtained by integrating $q(\vec{y}^{t} | \vec{Z}^{t-1})$ over $F(\vec{(y')}^{t})$, depends on the specific value of $\vec{(y')}^{t}$. In our examples below it can be evaluated exactly, but in some cases it may require estimation. 

Applying importance sampling on the probability space $\Omega^*$, and reintroducing the particle subscript $j$ gives
\begin{align*}
    E_{p^*}[l(\pi (\vec{(y')}^{t}, \vec{y}^{t})] 
    \approx \sum_{j=1}^n  w^{t}_j l((\vec{y}'_j)^{t})
\end{align*}
where the weights are given by:
\begin{eqnarray*}
    w^{t}_j &\propto& \frac{p^*(\vec{y}_j^{t}, (\vec{y}_j')^{t} | \vec{Z}^{t})} {q^*(\vec{y}_j^{t}, (\vec{y}_j')^{t} |\vec{Z}^{t-1})} \\
    &=& u(\vec{y}_j^{t} | (\vec{y}_j')^{t})\frac{p((\vec{y}'_j)^{t} | \vec{Z}^{t})}{q(\vec{y}_j^{t} | \vec{Z}^{t-1})} \\
    &=& \frac{u(\vec{y}_j^{t} | (\vec{y}_j')^{t})}{r(\vec{z}^{t} | \vec{Z}^{t-1})} \frac{q((\vec{y}_j')^{t} | \vec{Z}^{t-1})}{q(\vec{y}_j^{t} | \vec{Z}^{t-1})} \\
    & \propto & u(\vec{y}_j^{t} | (\vec{y}_j')^{t}) \prod_{i=1}^{t} w^{t}_{ij}
\end{eqnarray*}
with
\begin{eqnarray*}
w^{t}_{1j} &=& \frac{f((\vec{x}_j')^1)}{f(\vec{x}_j^1)} \frac{g((\vec{b}_j')^1 | (\vec{x}_j')^1)}{g(\vec{b}_j^1 | \vec{x}_j^1)}
\end{eqnarray*}
and
\begin{eqnarray*}
w^{t}_{ij} &=& \frac{f((\vec{x}_j')^i | (\vec{x}_j')^{i-1})}{f(\vec{x}_j^i | \vec{x}_j^{i-1})} \frac{g((\vec{b}_j')^i | (\vec{x}_j')^i, \vec{b}_j^{i-1})}{g(\vec{b}_j^i | \vec{x}_j^i, \vec{b}_j^{i-1})}
\end{eqnarray*}
for $i \geq 2$.
The term $r(\vec{z}^{t} | \vec{Z}^{t-1})$ is the same for all particles, and thus can be disregarded, since it will cancel after normalising the weights across particles. Note that $(\vec{b}_j')^i$ differs from $\vec{b}_j^i$ only for $i = t$, since the older observations $\vec{z}^{i}$ have already fully determined $\vec{b}_j^i$ for $i < t$. Also note that $w^{t}_{ij} = 1$ whenever both $(\vec{y}_j')^i = \vec{y}_j^i$ and $(\vec{y}_j')^{i-1} = \vec{y}_j^{i-1}$. Thus $w^{t}_{ij}$ only needs to be calculated for at most $2M$ values of $t$ for each $i$, specifically at the times when the coordinates $x^{i}_m$ and $x^{i-1}_m$ are observed for each $m$.

Using $u = u_1$, the normalised weights become:
\begin{eqnarray*}
w^{t}_j &=& \frac{\prod_{i=1}^t w^{t}_{ij}}{\sum_{j=1}^n \prod_{i=1}^t w^{t}_{ij}},
\end{eqnarray*}
(with zero-weight particles, which can arise as described above, discarded prior to normalisation, and $n$ reduced accordingly). Note that $u_1$ has the same value for all particles and hence cancels at normalisation.
For $u = u_2$, the normalised weights after cancelling terms are
\begin{eqnarray*}
w^{t}_j &=& \frac{\frac{1}{N_j} \prod_{i=1}^t v^{t}_{ij}}{\sum_{l=1}^n \frac{1}{N_l}\prod_{i=1}^t v^{t}_{il}}
\end{eqnarray*}
where $N_j$ is the normalisation constant obtained by integrating $q(\vec{y}^{t} | \vec{Z}^{t-1})$ over $F((\vec{y}_j')^{t})$, and the alternative partial weights are given by
\[
v^{t}_{1j} = w^{t}_{1j} f(\vec{x}_j^1)g(\vec{b}_j^1 | \vec{x}_j^1) = f((\vec{x}_j')^1)g((\vec{b}_j')^1 | (\vec{x}_j')^1)
\]
and
\begin{eqnarray*}
v^{t}_{ij} &=& w^{t}_{ij} f(\vec{x}_j^i | \vec{x}_j^{i-1}) g(\vec{b}_j^i | \vec{x}_j^i, \vec{b}_j^{i-1}) \\
&=& f((\vec{x}_j')^i | (\vec{x}_j')^{i-1}) g((\vec{b}_j')^i | (\vec{x}_j')^i, (\vec{b}_j')^{i-1})
\end{eqnarray*}
for $i \geq 2$.

\section{Simplifying the model when observations are missing at random}

The model in the previous section can be substantially simplified when observations are {\em missing at random}, by which we mean that $\vec{B}^t$ evolves independently of $\vec{X}^t$, that is 
\begin{align*}
    \vec{b}^1 | \vec{x}^1 &\sim g_1(\vec{b}^1 ) \mbox{ and } \\
    \vec{b}^t | \vec{x}^t, \vec{b}^{t-1} &\sim g_t(\vec{b}^t | \vec{b}^{t-1}).
\end{align*}
As a further simplification, in this section we limit our interest to expectations of the form:
\begin{align*}
    E_{p} \Big[l(\vec{X}^{t}) \Big] &=  \int_{\Omega_t'} l(\vec{X}^{t})\; p(\vec{X}^{t} | \vec{Z}^{t})\; \prod_{ \{ i \leq t  : z^t_{i} = - \} } \D x^i
\end{align*}
for functions $l: \Omega_t'  \rightarrow \Real$, where $\Omega_t' := \sigma^{-1}(\vec{Z}^{t})$ as before, remembering that $\vec{B}^{t}$ is fully determined by $\vec{Z}^{t}$, so that elements of $\Omega_t'$ all share the same value of $\vec{B}^{t}$. 

Under these assumptions, all $g$ terms cancel in the above expression for $p(\vec{X}^t, \vec{B}^t | \vec{Z}^t)$, giving $p(\vec{X}^t, \vec{B}^t | \vec{Z}^t) = p(\vec{X}^t | \vec{Z}^t)$. Moreover, all $g$ terms except the final term $g(\vec{b}^t | \vec{b}^{t-1})$ cancel in the above expression for $q(\vec{X}^t, \vec{B}^t | \vec{Z}^{t-1})$ and this last term may be removed by summing over $\vec{b}^t$ to obtain $q(\vec{X}^t | \vec{Z}^{t-1})$. In this case, it will be convenient to redefine $h$ and $r$ as follows:
\begin{eqnarray*}
h(\vec{X}^t) & := & f(\vec{x}^1)\prod_{i=2}^{t} f(\vec{x}^i | \vec{x}^{i-1})
\end{eqnarray*}
and
\begin{eqnarray*}
r(\vec{Z}^t) & := & \int_{\Omega_t'}  h(\vec{X}^t) \prod_{ \{ i \leq t, m \in \{1, \ldots, M \} : z^t_{im} = - \} } \D x^i_m
\end{eqnarray*}
with
\begin{eqnarray*}
r(\vec{z}^t | \vec{Z}^{t-1}) & := & \frac{r(\vec{Z}^t)}{r(\vec{Z}^{t-1})},
\end{eqnarray*}
and redefine $\Omega_t'$ as the projection of $\sigma^{-1}(\vec{Z}^t)$ onto the subspace obtained by discarding the second element of the pair $(\vec{X}^t,\vec{B}^t)$. With this modified notation, we can now write:
\[
q(\vec{X}^1) = f(\vec{x}^1),
\]
on $\Omega_1 = \Real^M$ and for $t \geq 2$,
\begin{eqnarray*}
    q(\vec{X}^t | \vec{Z}^{t-1}) & = & \frac{f(\vec{x}^1)}{r(\vec{z}^1)}  \prod_{i=2}^{t-1} \frac{f(\vec{x}^i | \vec{x}^{i-1}) }{r(\vec{z}^i | \vec{Z}^{i-1})} f(\vec{x}^t | \vec{x}^{t-1})
\end{eqnarray*}
on $\Omega_t = \Omega_{t-1}' \times \Real^M $ and 
\begin{eqnarray*}
    p(\vec{X}^t | \vec{Z}^t) 
& = & \frac{f(\vec{x}^1)}{r(\vec{z}^1)} \prod_{i=2}^t \frac{f(\vec{x}^i | \vec{x}^{i-1}) }{r(\vec{z}^i | \vec{Z}^{i-1})} 
\end{eqnarray*}
on $\Omega_t'$.

Sequential importance sampling in this context can be carried out using process similar to that described in the preceding section, but without simulating $\vec{B}^t$, and with $q(\vec{X}^t, \vec{B}^t | \vec{Z}^t)$ and $p(\vec{X}^t, \vec{B}^t | \vec{Z}^t)$ replaced by  $q(\vec{X}^t | \vec{Z}^t)$ and $p(\vec{X}^t | \vec{Z}^t)$ respectively. The above reasoning remains valid, with $\vec{y}^t$ and $(\vec{y}')^t$ replaced by $\vec{X}^t$ and $(\vec{X}')^t$ respectively. 

As before, there is flexibility in the choice of $u(\vec{X}^t | (\vec{X}')^t, \vec{Z}^t)$, which now is a distribution over the set $F((\vec{X}')^t, \vec{Z}^t)$ consisting of all possible $\vec{X}^t$ such that $\rho(\vec{X}^t, \vec{Z}^t) = (\vec{X}')^t$. One may set $u = u_1$, a uniform distribution over a hyper-rectangle contained in $F((\vec{X}')^t, \vec{Z}^t)$, chosen so that each undetermined coordinate of $\vec{X}^t$ is bounded independently of the others, leading to cancellation of the normalisation constant. The partial weights thus simplify to
\begin{eqnarray*}
w^{t}_{1j} &=& \frac{f((\vec{x}_j')^1)}{f(\vec{x}_j^1)}
\end{eqnarray*}
and
\begin{eqnarray*}
w^{t}_{ij} &=& \frac{f((\vec{x}_j')^i | (\vec{x}_j')^{i-1})}{f(\vec{x}_j^i | \vec{x}_j^{i-1})}
\end{eqnarray*}
for $i \in \{ 2, \ldots, t \}$. Note that since $\vec{B}^t$ is not simulated, it does not need to be corrected, and the $g$ ratio is not present in the partial weight $w^{t}_{tj}$.

Alternatively, one may set $u = u_2$, which now is the distribution $q(\vec{X}^t | \vec{Z}^{t-1})$ restricted to $F((\vec{X}')^t, \vec{Z}^t)$. In that case, the normalised weights after cancelling are
\begin{eqnarray*}
w^{t}_j &=& \frac{\frac{1}{N_j} \prod_{i=1}^t v^{t}_{ij}}{\sum_{l=1}^n \frac{1}{N_l}\prod_{i=1}^t v^{t}_{il}}
\end{eqnarray*}
where $N_j$ is the normalisation constant obtained by integrating $q(\vec{X}^{t} | \vec{Z}^{t-1})$ over $F((\vec{X}_j')^t, \vec{Z}^t)$. The alternative partial weights are given by
\[
v^{t}_{1j} = w^{t}_{1j} f(\vec{x}_j^1) = f((\vec{x}_j')^1)
\]
and
\[
v^{t}_{ij} = w^{t}_{ij} f(\vec{x}_j^i | \vec{x}_j^{i-1}) = f((\vec{x}_j')^i | (\vec{x}_j')^{i-1}) 
\]
for $i \geq 2$.

\section{Application to a stationary AR(1) model}
\label{sec:3}

Let us first introduce our importance sampling idea using a simple example. Consider a linear, normal and stationary AR(1) process for which
\begin{equation*}
    x^{t} = \varphi x^{t-1} + \eps^{t}
\end{equation*}
with $\eps^{t} \stackrel{iid}{\sim} \mathcal{N}(0, \sigma^{2})$. Then
\begin{equation*}
    (x^{t} | x^{t-1}) \sim \mathcal{N} (\varphi x^{t-1}, \sigma^{2}).
\end{equation*}
This process is stationary provided the initial distribution is
\begin{equation*}
    x^{1} \sim \mathcal{N} \Bigg (0, \frac{\sigma^2}{1- \varphi^2} \Bigg )
\end{equation*}
which is possible only if $|\varphi| < 1$. 

In this example, the dimension of the system is $M = 1$, and the trajectory of the system is $\vec {X}^{t} = (x^1, \dots, x^{t})$. Suppose that at time $t$, only some subset of the values $\vec {X}^{t} = (x^1, \dots, x^{t})$ have so far been observed. However, those that have been observed are known without error. Note that because $M=1$, $\vec{z}^{t}$ may be represented as a row vector with $t$ coordinates, and $\vec{Z}^{t}$ may be represented as a lower triangular $t \times t$ matrix, in which row $i$ contains the vector $\vec{z}^{i}$, padded with zeroes at the right. Similarly, $\vec{b}^{t}$ can be represented as a row vector with $t$ coordinates and $\vec{B}^{t}$ can be represented as a lower triangular $t \times t$ binary matrix in which row $i$ is the vector $\vec{b}^{i}$.

We model $b_i^{t}$, when $b_i^{t-1} = 0$, as Bernoulli with probability $\theta$:
\begin{equation*}
    (b_i^{t} | b_i^{t-1} = 0) \sim Bernoulli(\theta).
\end{equation*}
Thus $\vec{B}^{t}$ evolves independently of $\vec{X}^{t}$, so that the conditional distribution of $\vec{b}^{t}$ given $(x^{t}, \vec{b}^{t-1})$ is of the form $g(\vec{b}^{t} | x^{t}, \vec{b}^{t-1} ) = g( \vec{b}^{t} | \vec{b}^{t-1} )$. As we saw above, in this case the sequential importance sampling weights do not depend on $\vec{B}^t$.

\subsection{Sequential importance sampling for an AR(1) model}
\label{sec:5}

We first use the sequential importance strategy described in Section~\ref{sec:2} with $u = u_1$. At time $t$, the prior density $q(\vec{X}^{t} | \vec{Z}^{t-1})$ is defined on the subspace $\Omega_t = \prod_{i=1}^t \mathbb{X}_i^{t-1} \subseteq \mathbb{R}^t$ with $\mathbb{X}_i^{t-1} = \mathbb{R}$ for $z_i^{t-1} = -$, and $\mathbb{X}_i^{t-1} = \{ z_i^{t-1} \}$ otherwise. That is, $\Omega_t$ is the subspace of $\mathbb{R}^t$ corresponding to observed co-ordinates being fixed. Note we adopt the convention $z_t^{t-1} = -$, since the system state $x^t$ is not observed at time $t-1$.  The posterior density $p(\vec{X}^{t} | \vec{Z}^{t})$ is defined on the subspace $\Omega_t' = \prod_{i=1}^t \mathbb{X}_i^{t} \subseteq \mathbb{R}^t$ with $\mathbb{X}_i^{t} = \mathbb{R}$ for $z_i^{t} = -$ and $\mathbb{X}_i^{t} = \{ z_i^{t}\}$ otherwise. 

For each particle $j$, the partial weights at time $t$ are calculated with the following formula
\begin{align} \label{eq:w1}
    w_{1j}^t &= \frac{\exp \left[ -\frac{1 - \varphi^2}{2 \sigma^2}{((x')^1)^2} \right]}{\exp \left[ -\frac{1 - \varphi^2}{2 \sigma^2}{(x^1)^2} \right]} \mbox{ and } \\ 
    w_{ij}^t &= \frac{\exp \left[ -\frac{1}{2 \sigma^2}{((x')^i - \varphi (x')^{i-1})^2} \right]}{\exp \left[ -\frac{1}{2 \sigma^2}{(x^i - \varphi x^{i-1})^2} \right]} \label{eq:w2}
\end{align}
for $i \in \{ 2, \ldots, t-1 \}$.

Our algorithm outputs a set of particles, each with an associated weight calculated as explained above, then performs a resampling step which draws from the generated particle set with probability proportional to the weights.

We have compared the distributions for the missing observations to distributions obtained analytically as explained in the section below.

\subsection{Gold Standard: Analytical estimation of missing observations for an AR(1) model when the available observations are exact}
\label{sec:6}

Here we derive analytical expressions for posterior distributions of missing values in the AR(1) context. It will be convenient to think of the missing values as organised into blocks, where a `block' is a sequence of one or more contiguous missing values.

The Kalman Smoother is not well suited to exact observations, since the one-step ahead error (for a definition see \cite{Young}) cannot be defined. However, we can use the model definition and knowledge of the observations before and after the missing terms to estimate the posterior for those missing terms analytically.

Suppose at time $t$, the system state at earlier times $\tau$ and $m$ is known, where $1 \leq \tau < m \leq t$, but the system state at intermediate times is unobserved. That is, $z^t_{\tau} = x^{\tau}$, $z^t_{m} = x^m$, but $z^t_i = -$ for $i \in \{ \tau+1, \ldots, m-1\}$. Since the AR(1) model is Markovian, any knowledge we may have about the system state prior to time $\tau$ or after time $m$ is irrelevant, given $x^{\tau}$ and $x^{m}$. Thus the posterior distribution for the system state at times $i \in \{ \tau+1, \ldots, m-1\}$ is given by

\begin{align*}
   p(x^i | \vec{z}^t)  &= p(x^i | x^{\tau}, x^m) \\
    &= \frac{p(x^m | x^i) p(x^i | x^{\tau})}{p(x^m | x^{\tau})} \\
\end{align*}
For a Gaussian AR(1) model, we have 

\[
p(x^i | x^{\tau}) \propto \exp\Big(-\frac{1}{2} R_{\tau,i}\Big)
\]
where

\[
    R_{\tau,i} = \frac{(x^i - \varphi^{i-\tau} x^{\tau})^2}{\sigma^2 \sum_{j=0}^{i-\tau-1} \varphi^{2j}}.
\]
Similarly,

\begin{align*}
    p(x^m | x^i) &\propto \exp\Big(-\frac{1}{2} R_{i,m}\Big) \mbox{ and } \\
    p(x^m | x^{\tau}) &\propto \exp\Big(-\frac{1}{2} R_{\tau,m}\Big)
\end{align*}
where $R_{i,m}$ and $R_{\tau,m}$ are given by similar expressions to that for $R_{\tau,i}$. Putting this together and normalising gives 

\begin{align*}
   p(x^i | \vec{z}^t)  &= c \exp\Big(-\frac{1}{2} R\Big)
\end{align*}
where $c$ is a normalisation constant and

\begin{align*}
    R &= \frac{(x^m - \varphi^{m-i} x^i)^2}{\sigma^2 \sum_{j=0}^{m-i-1} \varphi^{2j}} + \frac{(x^i - \varphi^{i-\tau} x^\tau)^2}{\sigma^2 \sum_{j=0}^{i-\tau-1} \varphi^{2j}} - \\
    &\frac{(x^m - \varphi^{m -\tau} x^\tau)^2}{\sigma^2 \sum_{j=0}^{m-\tau-1} \varphi^{2j}}
\end{align*}
This expression factorizes to give

\begin{align*}
    &R = \frac{\sum_{j=0}^{m-\tau - 1} \varphi^{2j}}{\sigma^2 \sum_{j=0}^{m-i-1} \varphi^{2j} \sum_{j=0}^{i-\tau-1} \varphi^{2j}} \\
    &\Bigg\{ x^i - \Bigg[\frac{\varphi^{m-i} \sum_{j=0}^{i-\tau-1} \varphi^{2j} x^{m} + \varphi^{i - \tau} \sum_{j=0}^{m - i - 1} \varphi^{2j} x^{\tau}}{\sum_{j=0}^{m-\tau-1} \varphi^{2j}}\Bigg] \Bigg\}^2
\end{align*}
therefore

\begin{align*}
      &p(x^i | x^{\tau}, x^{m}) = \\
      &\mathcal{N} \Bigg(\frac{\varphi^{m-i} \sum_{j=0}^{i-\tau-1} \varphi^{2j} x^{m} + \varphi^{i-\tau} \sum_{j=0}^{m-i-1} \varphi^{2j} x^{\tau}}{\sum_{j=0}^{m-\tau -1} \varphi^{2j}}, \\ &\frac{\sigma^2 \sum_{j=0}^{m-i-1} \varphi^{2j} \sum_{j=0}^{i-\tau - 1} \varphi^{2j}}{\sum_{j=0}^{m-\tau - 1} \varphi^{2j}} \Bigg).
\end{align*}
We have run our simulations with 1,000 particles and 30 times and compared our methodology with this analytical solution showing that our algorithm and the gold standard agree well (see Figure 1).

\section{Application to an invasive species model}
\label{sec:7}

This second example illustrates the applicability of our method to invasive species models. For simplicity, we consider a one-dimensional environment such as a river without tributaries or confluences. We divide the length of the river into $M$ sections or cells and suppose the first introduction of the invasive species occurs in a cell with index $\mu$. From this cell, the invasion can propagate only in the immediately adjacent sections of the river, namely the cells indexed by $\mu-1$ and $\mu+1$ (an unrealistic assumption for most water-borne invasive species, but convenient for illustrative purposes). At each time step, a cell becomes infested from an adjacent infested cell with probability $\theta$. We represent the state of the system at time $t \in \{ 1, \dots, T \}$ by a binary vector $\vec{x}^{t} = (x_1^{t}, \dots, x_{M}^{t})$. These vectors indicate the state of each cell at time $t$: $x_m^t = 1$ if cell $m$ is infested and $x_m^t = 0$ otherwise. The invasion will be complete, (that is, all cells will be infested) at an unknown time $T$. 
The invasion expands in two directions (left and right) simultaneously with the same probability $\theta$ of expansion by one cell applicable in both directions. It is convenient to collect these vectors to form a $t\times M$ matrix $\vec{X}^{t}$, in which each row $i$ is the binary vector $\vec{x}^{i}$.

We store information regarding which cells have been observed to contain the invasive species at or before time $t$ in a binary matrix $\vec{b}^t = (b_{im}^t)$, where $b_{im}^t = 1$ if the state of cell $m$ at time $i$ is known by time $t$, and $b_{im}^t = 0$ otherwise. The observations available at time $t$ are represented by a matrix $\vec{z}^t = (z_{im}^t)$, where we set $z_{im}^t = x^i_m$ if $b_{im}^t = 1$ and $z_{im}^t = -$ otherwise. However, in this example only presences may be observed, so $b_{im}^t = 1$ implies $x_m^i = 1$ and thus $z_{im}^t = 1$. Thus $\vec{b}^t$ and $\vec{z}^t$ fully determine each other: $z_{im}^t = 1$ if and only if $b_{im}^t = 1$, and $z_{im}^t = -$ if and only if $b_{im}^t = 0$. Moreover, $\vec{b}^t$ fully determines $\vec{b}^{t'}$ at any earlier time $t' < t$ by discarding the matrix rows corresponding to times after $t'$, and similarly for $\vec{z}^t$. Thus it is not necessary to form $\vec{B}^t = (\vec{b}^1,\ldots,\vec{b}^t)$ or $\vec{Z}^t = (\vec{z}^1,\ldots,\vec{z}^t)$ as in the general case: the same information is already contained in $\vec{b}^t$ or $\vec{z}^t$.

We will model the new observations on the left and on the right of the previously observed nests in the following way. We will consider observations made by a probe that will check only those unobserved cells adjacent to cells where the invasive species has previously been detected. For example, if a cell is observed as infested at time $t-1$, at time $t$ the adjacent cell in which no invader has previously been detected will be probed. If that cell is found to be infested, adjacent cells will also be probed, and so on until a cell in which no invaders are detected is found. The probe might fail to observe an invader that is present, and hence searching may stop without finding all invaded cells. The probability of observing invaders in an infested cell is denoted $\varphi$. 

In this example, as for the AR(1) model, corrections are deterministic. That is, values of $\vec{X}^t$ and $\vec{b}^t$ simulated prior to receiving the new observations $\vec{z}^t$, will be corrected to the values $(\vec{X}')^t$ and $(\vec{b}')^t$ implied by $\vec{z}^t$ when those observations become available.

\subsection{Bayesian inference in a partially observed state space}
\label{sec:8}

We want to determine the posterior distribution $p(\vec{X}^{t}, \vec{b}^{t} | \vec{z}^{t})$ and this is obtained, as for the AR(1) model, by renormalising the prior at time $t$, $q(\vec{X}^{t}, \vec{b}^{t} | \vec{z}^{t-1})$, over values consistent with the observations $\vec{z}^{t}$

\begin{equation*}
    p(\vec{X}^{t}, \vec{b}^{t} | \vec{z}^{t})  =  \frac{q(\vec{X}^{t}, \vec{b}^{t} | \vec{z}^{t-1})}{r(\vec{z}^{t} | \vec{z}^{t-1})}
\end{equation*}
where 
\begin{equation*}
    r(\vec{z}^{t} | \vec{z}^{t-1}) = \sum_A q(\vec{X}^{t}, \vec{b}^{t} | \vec{z}^{t-1})
\end{equation*}
with $A = \{ \vec{X}^t \in 2^{t \times M} | x^i_m = 1 \mbox{ whenever } z_{im}^t = 1 \}$.

The prior for time $t$, $q(\vec{X}^{t}, \vec{b}^t | \vec{z}^{t-1})$ represents our knowledge about $(\vec{X}^{t}, \vec{b}^t)$ before we observe $\vec{z}^{t}$ and can be written as
\begin{align}
    & q(\vec{X}^{t}, \vec{b}^{t} | \vec{z}^{t-1}) = g(\vec{b}^{t} | \vec{x}^{t}, \vec{b}^{t-1}) f(\vec{x}^{t} | \vec{x}^{t-1}) p(\vec{X}^{t-1}, \vec{b}^{t-1} | \vec{Z}^{t-1}). \label{eq:3}
\end{align}
The distribution for the observations is:
\begin{align*}
    & g(\vec{b}^{t} | \vec{x}^{t}, \vec{b}^{t-1}) = g_R(\vec{b}^{t} | \vec{x}^{t}, \vec{b}^{t-1}) g_L(\vec{b}^{t} | \vec{x}^{t}, \vec{b}^{t-1})
\end{align*}
where $g_R(\vec{b}^{t} | \vec{x}^{t}, \vec{b}^{t-1})$ is the probability of detecting invaders on the right of the cells previously observed as infested and $g_L(\vec{b}^{t} | \vec{x}^{t}, \vec{b}^{t-1})$ is the probability of detecting invaders on the left of the cells previously observed as infested. These two events are independent.

Let $a_{t}$ be the largest integer in the set $\{ m : b_{tm}^{t} = 1 \}$, that is, the index of the highest numbered cell at which the invader has been observed at time $t$ on the right of the invasion. Then the number of new probes at which the invader is successfully detected at time $t$ is $a_{t} - a_{t-1}$. Probing on the right can stop in two ways: either because all invaded sites on the right have been found, in which case $g_R(\vec{b}^{t} | \vec{x}^{t}, \vec{b}^{t-1}) = \varphi^{a_{t} - a_{t-1}}$, or because the last probe on the right failed to detect an invaded cell, in which case $g_R(\vec{b}^{t} | \vec{x}^{t}, \vec{b}^{t-1}) = \varphi^{a_{t} - a_{t-1}} (1 - \varphi)$.

Similarly, let $c_{t}$ be the smallest integer in the set $\{ m : b_{tm}^{t} = 1 \}$, that is, the index of the smallest cell at which the invader has been observed at time $t$ on the left of the invasion. Then the number of new probes at which the invader is successfully detected at time $t$ is $c_{t} - c_{t-1}$.

The distribution for the observation will therefore be:
\begin{align*}
    g(\vec{b}^{t} | \vec{x}^{t}, \vec{b}^{t-1}) = &\varphi^{(a_{t} - a_{t-1})} \bigg(1 - \varphi \bigg)^{\one {a_{t} \neq \gamma_{t}}}\\
    &\varphi^{(c_{t} - c_{t-1})} \bigg( 1 - \varphi \bigg)^{\one {c_{t} \neq \beta_{t}}}
\end{align*}
were $\gamma_{t}$ is the limit of the invasion on the right and $\beta_{t}$ is the limit of the invasion on the left. Notice that for simplicity we will assume that we observe the beginning of the invasion at time $t=1$. 

The distribution $f(\vec{x}^{t} | \vec{x}^{t-1})$ if the expansion can happen in two directions is:
\begin{equation*}
    f(\vec{x}^{t} | \vec{x}^{t-1}) = f_R(\vec{x}^{t} | \vec{x}^{t-1}) f_L(\vec{x}^{t} | \vec{x}^{t-1}) 
\end{equation*}
separating the expansion on the left and on the right of the invasion, since
these two expansions are independent. Then, $f_R(\vec{x}^{t} | \vec{x}^{t-1}) = \theta$ if the invasion has expanded one cell to the right, and $f_R(\vec{x}^{t} | \vec{x}^{t-1}) = 1 - \theta$ if no expansion to the right has occurred, and $f_R(\vec{x}^{t} | \vec{x}^{t-1}) = 0$ for any other $\vec{x}^{t}$. The same applies to the expansion on the left. Therefore,
\begin{align*}
    f(\vec{x}^{t} | \vec{x}^{t-1}) &= f_R(\vec{x}^{t} | \vec{x}^{t-1}) f_L(\vec{x}^{t} | \vec{x}^{t-1})\\
    & = \theta^{k} (1-\theta)^{1-k}\theta^{h} (1-\theta)^{1-h} 
\end{align*}
with $k \in \{0,1\}$ and $h \in \{0,1\}$, where $k = 1$ if the invasion expanded one cell to the right and $k = 0$ otherwise, and $h = 1$ if the invasion expanded one cell to the left and $h = 0$ otherwise.

Substituting and expanding the recursion in equation \eqref{eq:3} we have
\begin{align}
    & q(\vec{X}^{t},\vec{b}^{t} | \vec{b}^{t-1}) =  \prod_{i=2}^{t} \varphi^{(a_{i} - a_{i-1})}\bigg(1 - \varphi \bigg)^{\one {a_i \neq \gamma_i}} \nonumber \\
    & \varphi^{(c_{i} - c_{i-1})} \bigg(1 - \varphi\bigg)^{\one {c_i \neq \beta_i}} \; \prod_{i=2}^{\min\{t,r\}} \theta^{k_i} (1-\theta)^{1-k_i}  \label{eq:4} \\ 
    & \prod_{i=2}^{\min\{t,l\}} \theta^{h_i} (1-\theta)^{1-h_i} \nonumber
\end{align}
where $l$ is the time at which the invasion has reached the left end of the modelled region, $r$ is the time at which the invasion has reached the right end of the modelled region, $k_i \in \{0,1\}$ and $h_i \in \{0,1\}$ where $k_i = 1$  if the invasion expanded one cell to the right at time $i$ and $k_i = 0$ otherwise, and $h_i = 1$ if the invasion expanded one cell to the left at time $i$ and $h_i = 0$ otherwise.

For each particle j, the partial weights at time t are calculated with the following formula

\begin{equation*}
    w_{1j}^t = 1
\end{equation*}

\begin{align*}
    w_{ij}^t = & \frac{\varphi^{a'_i} \Big(1-\varphi \Big)^{\one {a'_i \neq \gamma_i}} \varphi^{c'_i}(1-\varphi)^{1-\one {c'_i = \beta_i}}}{\varphi^{a_i} \Big(1-\varphi \Big)^{\one {a_i \neq \gamma_i}} \varphi^{c_i} \Big(1-\varphi \Big)^{1-\one {c_i = \beta_i}}} \\ \nonumber
    & \Bigg(\frac{\theta^{k'_i} (1-\theta )^{1-k'_i} }{ \theta^{k_i} (1-\theta )^{1-k_i}} \Bigg)^{\one {i \leq min\{ t, r\}} } \\ \nonumber
    & \Bigg( \frac{ \theta^{('_i} (1-\theta )^{1-h'_i} }{ \theta^{h_i} (1-\theta )^{1-h_i} } \Bigg)^{\one {i \leq min\{ t, l\}}}.
\end{align*}

Figure 2 shows a series of heat maps for 3 possible invasions and observations simulated from the distributions in the river example. The map is a visualization of the probability of finding an invader in each location at each time for different values of the observation parameter $\varphi$ but for the same value of the expansion  parameter $\theta$. The length of the vertical axis represents the time at which the simulated invasion reached an end and all cells are occupied: for example in figure A1 at time 165 in the simulated invasion all the cells are invaded. The image shows how we get a more accurate representation of the invasion with a higher number of observations. We can also see that the algorithm becomes less accurate when we have only few observations.

\begin{figure*}
    \subfloat[fig 1]{\includegraphics[width=0.5\linewidth]{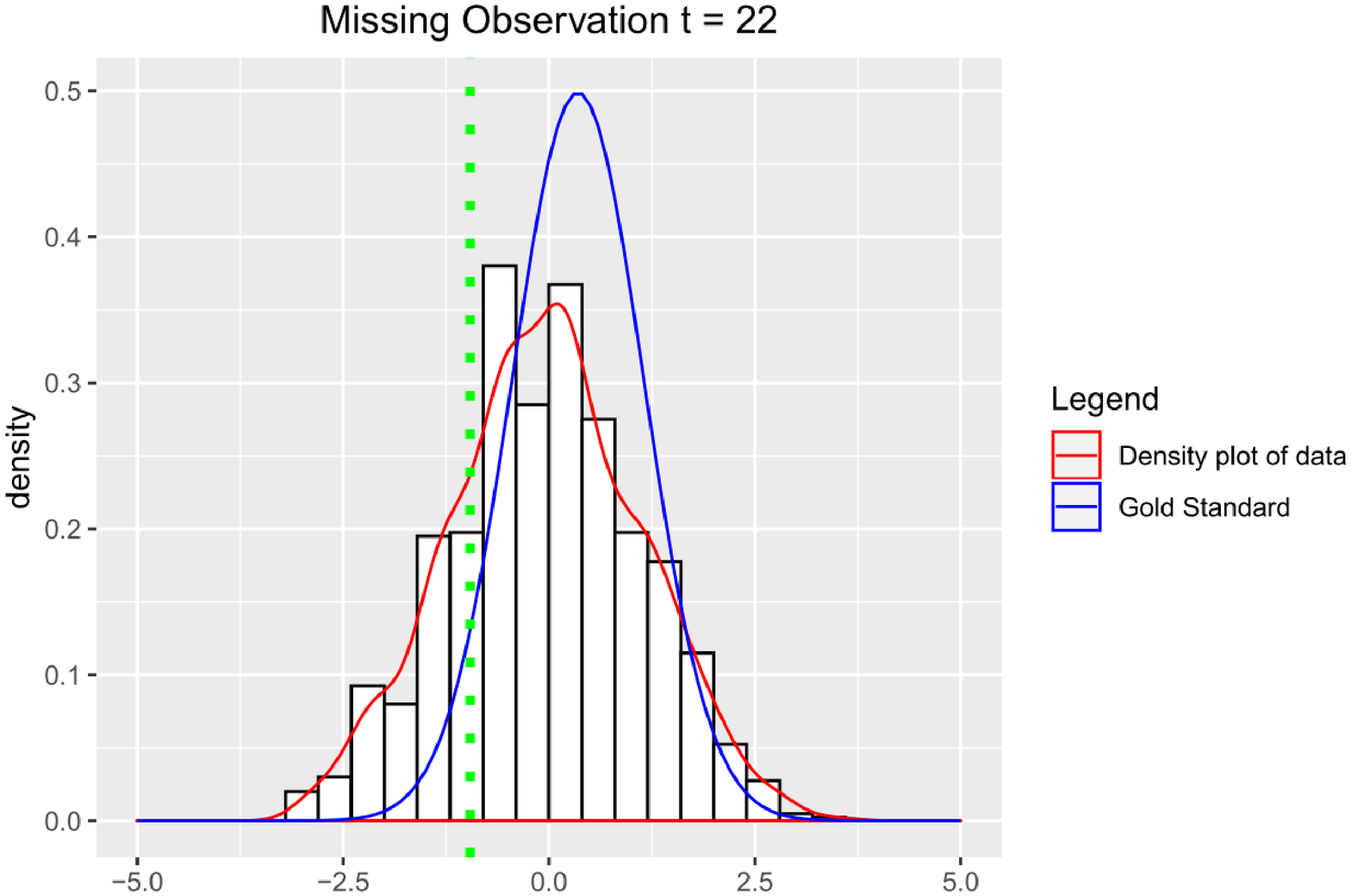}}
    \subfloat[fig 2]{\includegraphics[width=0.5\linewidth]{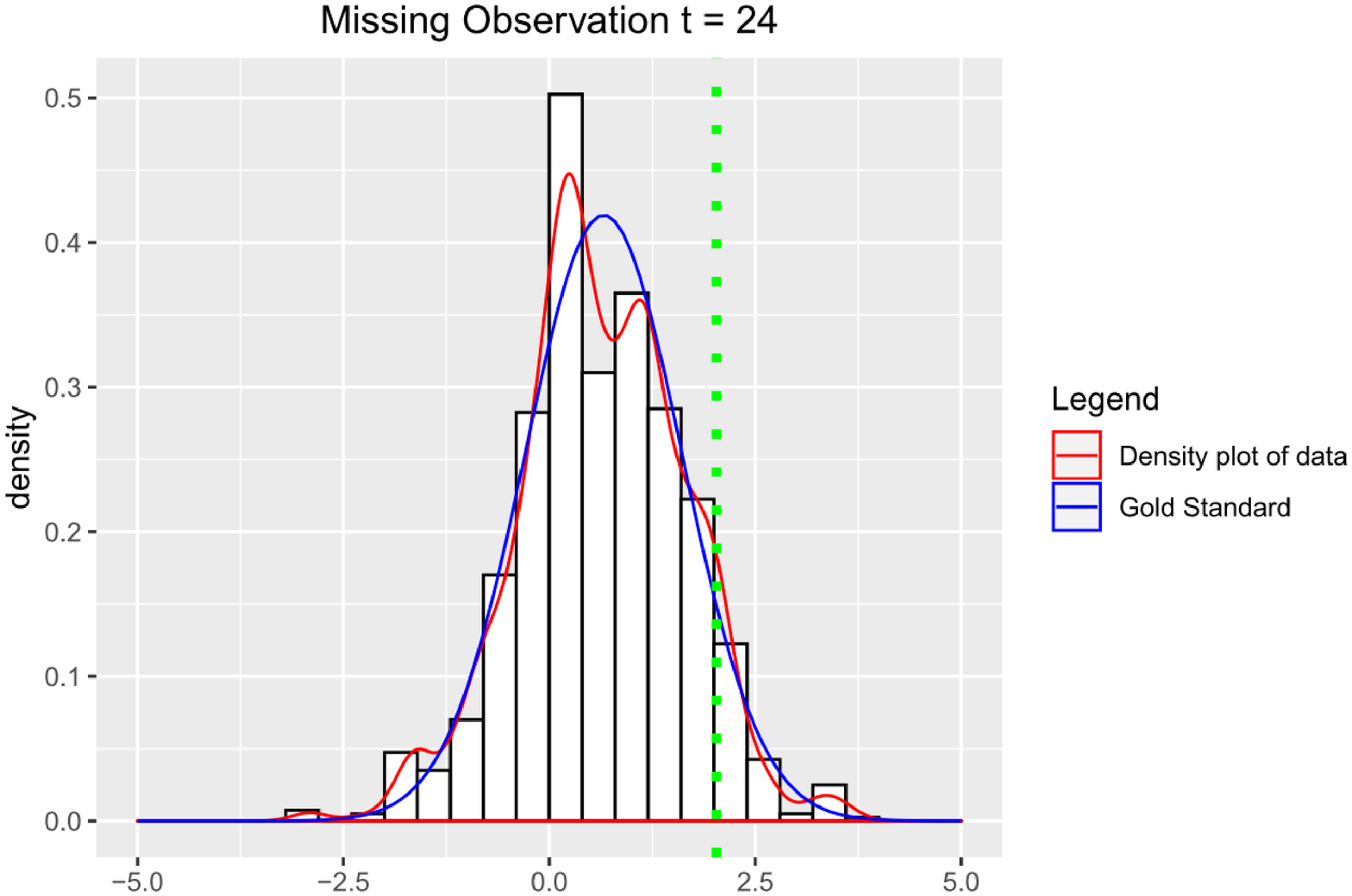}}\\
    \subfloat[fig 3]{\includegraphics[width=0.5\linewidth]{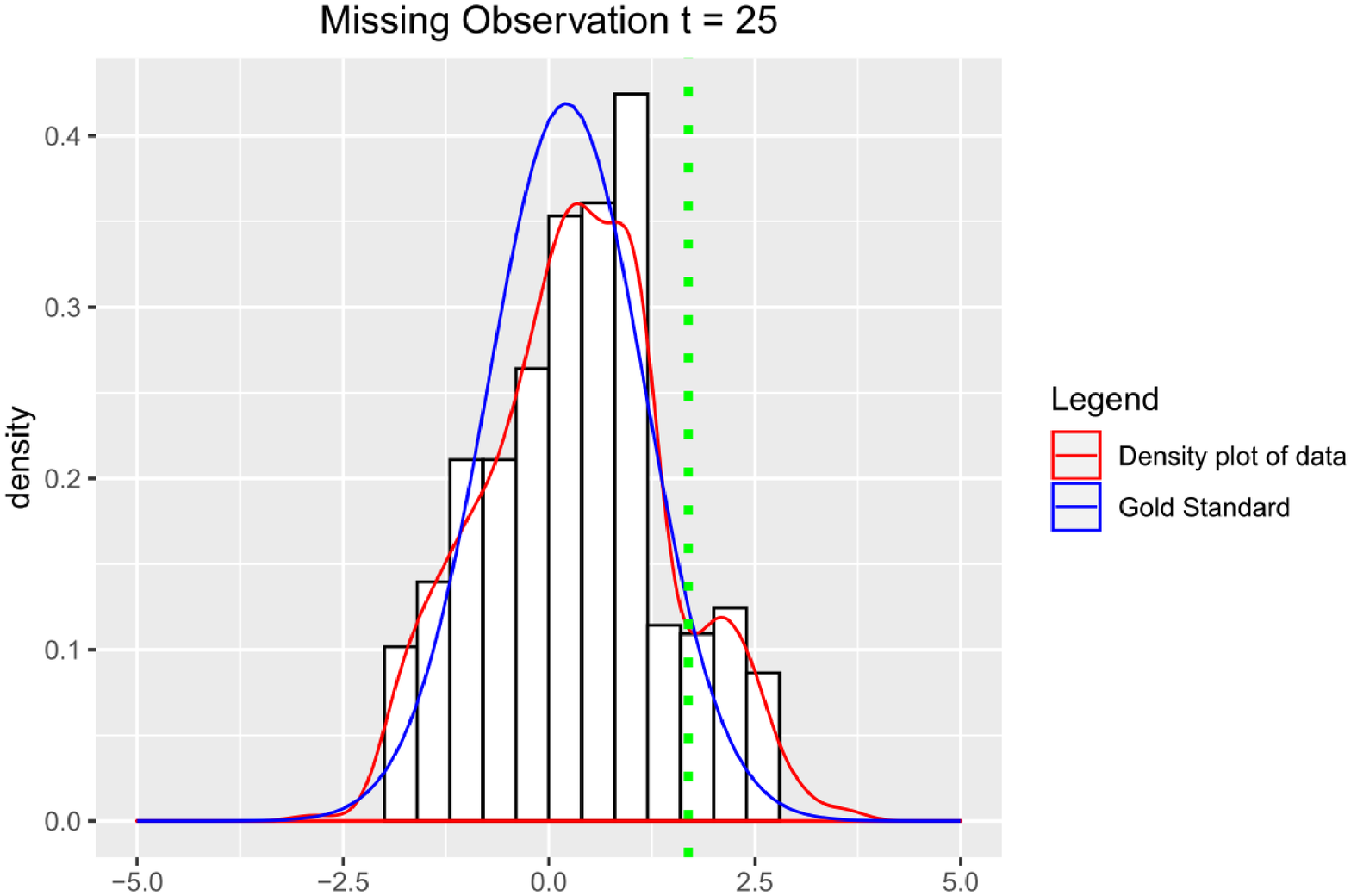}}
    \subfloat[fig 4]{\includegraphics[width=0.5\linewidth]{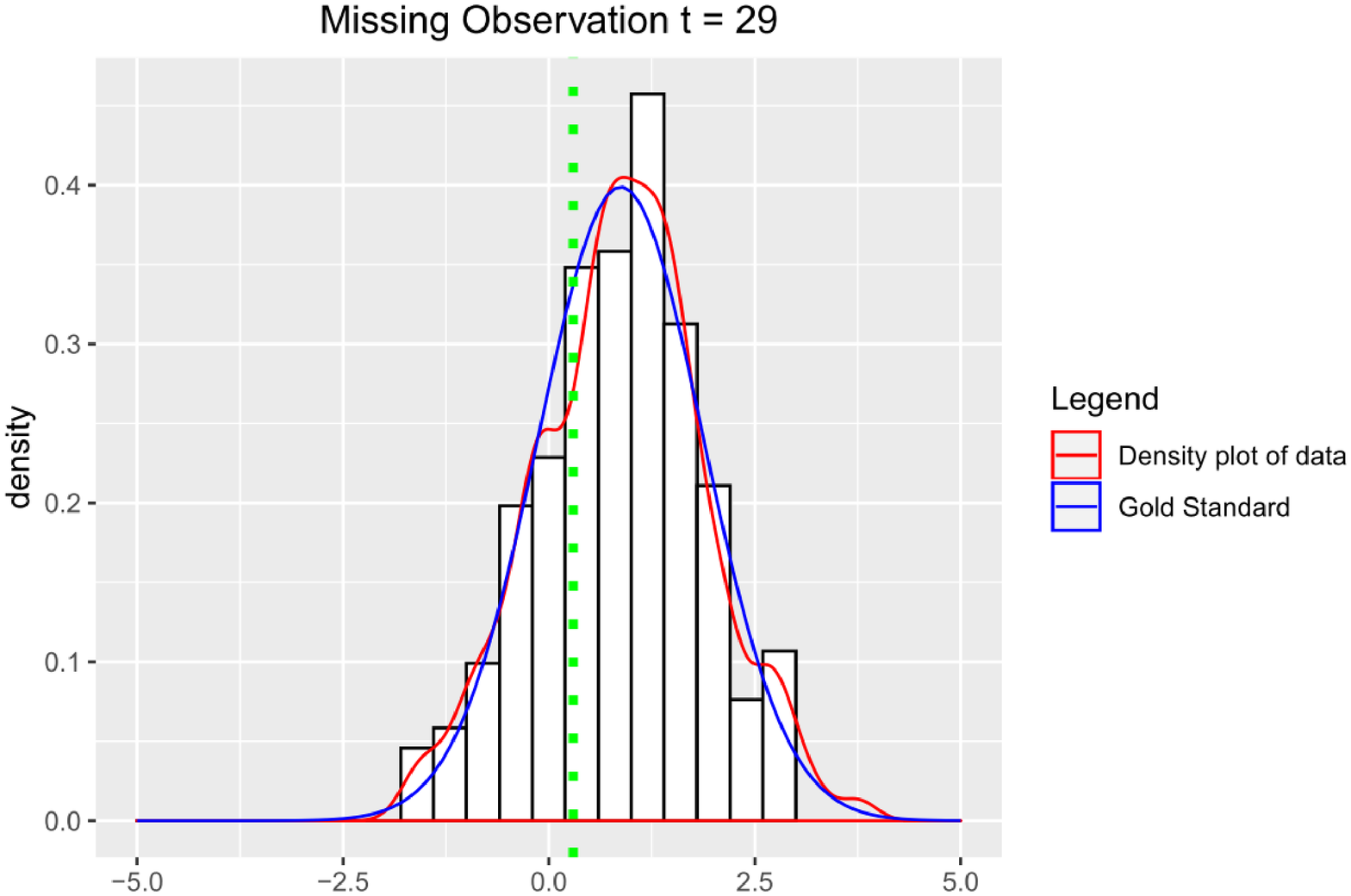}}\\
    \subfloat[fig 5]{\includegraphics[width=0.5\linewidth]{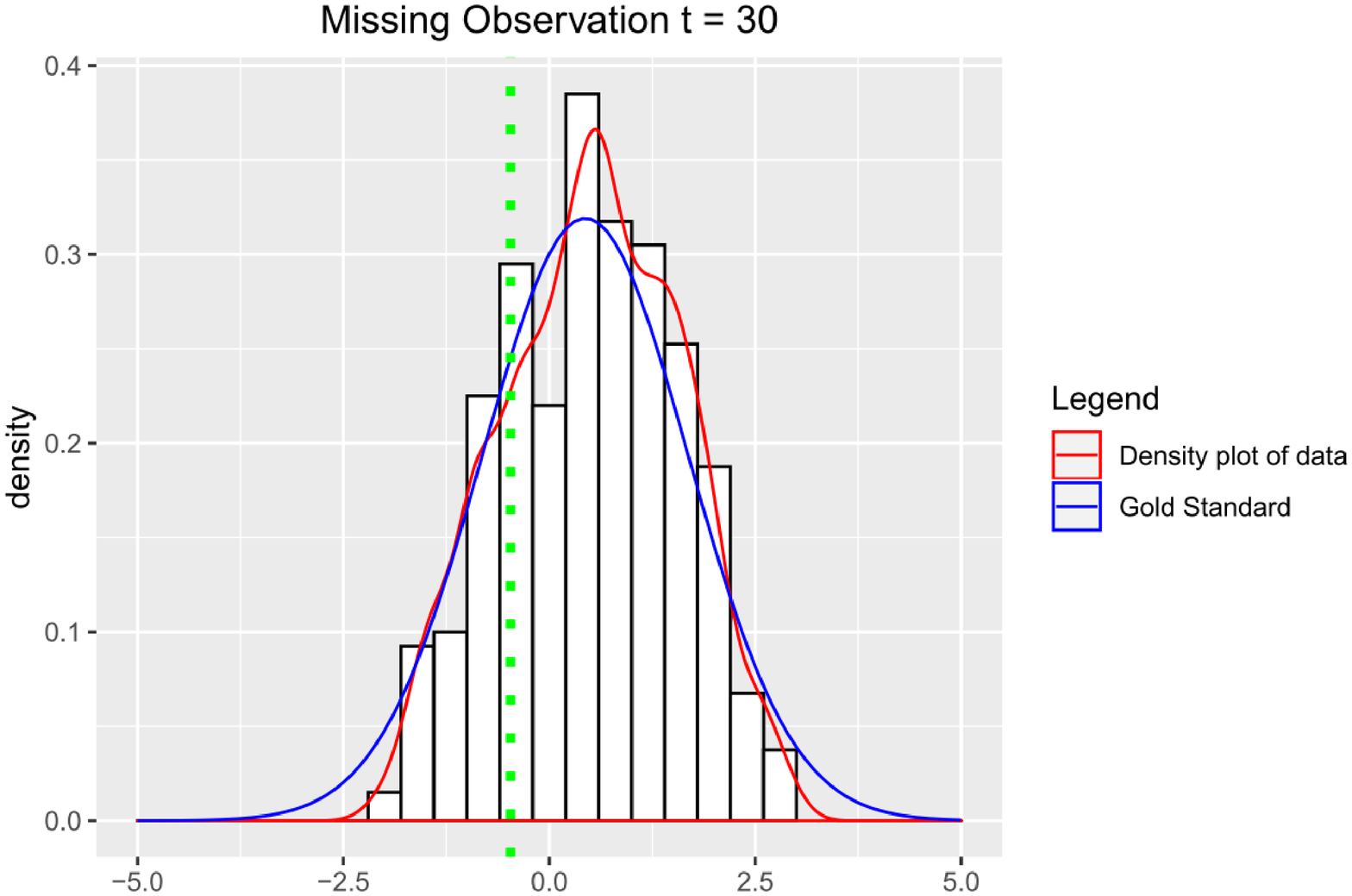}}
    \caption{AR(1) MODEL: Simulations with 1,000 particles and 30 times compared to the gold standard for the 5 missing times of the AR(1) model. In Green the value for the observation. The AR(1) model had parameter $\varphi = 0.5$, variance $\sigma^2 = 1$. The Bernoulli distribution had parameter $p = 0.2$}
\label{fig:1}
\end{figure*}

\begin{figure*}
    \includegraphics[width=\textwidth]{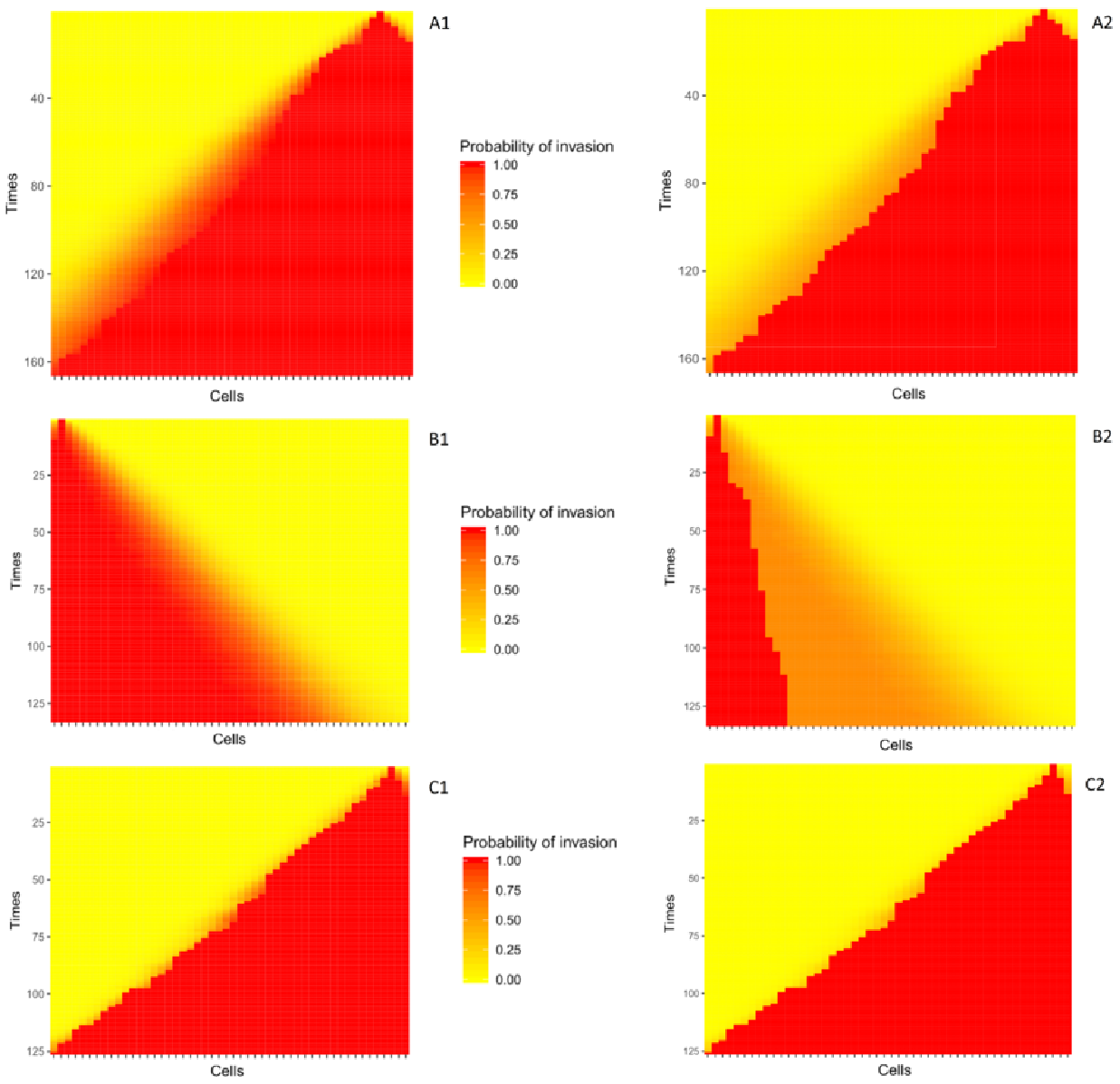}
    \caption{RIVER INVASION: Simulations with 1,000 particles and 50 cells of 3 possible river invasions all with probability of invasion $\theta = 0.3$ but with different values of the parameter $\varphi$ for the probability of the observations. Figures A1, B1 and C1 show the simulations of the 3 invasions while figures A2, B2 and C2 show the same simulations but with the observations superimposed in red. A1 and A2 show the invasion with $\varphi = 0.3$, B1 and B2 show the invasion with $\varphi = 0.1$. Figure C1 and C2 show the invasion with $\varphi = 0.8$.}
    \label{fig:2}
\end{figure*}

\section{Conclusions and further developments}
\label{sec:9}

The present investigation shows how our new methodology can be used to estimate the past and present state of a system for which incomplete information is continuously gathered. In particular in the case of invasive species it will be well suited to estimate the past trajectories and current extent of an invasion. This could then be used to help understand the efficacy of an eradication program, or to inform management actions regarding where to search for currently hidden invaders, or to assess the risk from currently unobserved individuals.

The results for the AR(1) model compared to the gold standard show that our simulations closely approximate analytical solutions, while the river example shows the applicability of the method to a simple invasive species problem.

As mentioned in the introduction, different methods exist to numerically solve estimation problems with exact but incomplete observations in an online manner. However, our method is the first one, to the best of our knowledge, that corrects previous estimations as each new observation is acquired.

The methodology could also be used in the more complex scenario where not only we have entire cases unobserved, but also the number of missing cases is unknown, for example, in predictive modeling of a species' geographical distribution. In this context, we are planning to apply our new method to a specific invasive species problem: the fire ants invasion in Queensland, Australia. This application will seek to improve and complement the existing agent based approach developed by Keith and Spring (\cite{Keith}). Their method consisted of constructing a likelihood model in terms of some unknown parameters that included, among other things, the phylogeny, jump type, founding type and treatment success rate. The posterior distribution was then sampled using a generalised Gibbs technique that enabled trans-dimensional sampling, which is used when the number of parameters is unknown, as in their case. However, their method needed to be re-run each time new data was available and could not be updated online.

In this new application, the methodology will need to be adapted to continuous time models. We also envisage the possibility of improving our approach to allow for non-deterministic revision of imputed values that, while not inconsistent with later observations, can nevertheless be updated in light of new information. This will require a different definition of the probability $p^*$ and some more theoretical proof of its existence and validity. 

Our algorithm is reasonably fast, as the weights involve cancellations that significantly reduce the calculations required. However, we might need to optimise and further maximise efficiency when running with bigger data sets. For example, we could apply a more efficient resampling step.

\clearpage
\appendix

\begin{algorithm}[H]
\caption{SIR with corrections for an AR(1) model}\label{SIR_AR1}
 \begin{algorithmic}

 \State  \bf{Initialize:} \normalfont At time t = 1
            
\begin{enumerate}
	\item For $j = 1, \dots , n$
	\begin{enumerate}
		\item Sample $(x_{1})_j \sim \mathcal{N} \Bigg (0, \frac{\sigma^2}{1- \varphi^2} \Bigg)$ and $(b_1)_j \sim Bernoulli(\theta)$
		\item Evaluate the importance weights up to a normalising constant:
		\[
		\tilde{w}^{1}_{j} = 1
		\]
	\end{enumerate}
	\item For $j = 1, \dots , n$ normalise the importance weights: 
	\[
	w^{1}_{j} = \frac{1}{n}
	\]
\end{enumerate}

 \State  \bf{Iterate:} \normalfont For $t$ from 2 to $N$

\begin{enumerate}
	\item For $j = 1, \dots , n$
	\begin{enumerate}
  		\item sample $(x_t)_j \sim \mathcal{N} (\varphi (x_{t-1})_j, \sigma^{2})$ and $b^t_{j} \sim Bernoulli(\theta)$.
		\item {Correct every element of $\vec{x}^t_j$ in light of the data $\vec{z}^t$ to obtain new samples $(\vec{x'})^t_j$ directly substituting the new data: for $s = 1, \dots ,t$ if $(z^t)_s \neq -$, $((x')_j^t)_s = (z^t)_s$ else if $(z^t)_s = -$, $((x')_j^t)_s = (x^t_j)_s$.}
		\item Correct every element of $\vec{b}_i^t$ in light of the data $\vec{z}^t$ to obtain new samples $(\vec{b'})^t_j$ substituting 0s when $(z^t)_s = -$ and 1s otherwise for $s = 1, \dots, t$.
		\item Evaluate the weights up to a normalising constant using equations (\ref{eq:w1}) (\ref{eq:w2}):
		\[
		\tilde{w}^{t}_{j} = \frac{q(\vec{(x')}^{t}_j, \vec{(B')}^{t}_j | \vec{Z}^{t-1})}{q(\vec{x}^{t}_j, \vec{B}^{t}_j | \vec{Z}^{t-1})}
		\]
	\end{enumerate}
	\item For $j = 1, \dots , n$ normalise the importance weights:
	\[
	w^{t}_{j} = \frac{\tilde{w}^t_j}{\sum_{k=1}^{n}\tilde{w}^{t}_k}
	\]
	\item Perform resampling
	\begin{enumerate}
	    \item Draw $n$ particles from the current particle set with probabilities proportional to their weights. Replace the current particle set with the new $n$ particles.
	    \item Set $w^t_j=1/n$.
	\end{enumerate}
\end{enumerate}
  
 \end{algorithmic}
\end{algorithm}

\begin{algorithm}[H]
\caption{SIR with corrections for an river invasion}\label{SIR_River}
 \begin{algorithmic}

\State  \bf{Initialize:} \normalfont At time t = 1
            
\begin{enumerate}
	\item For $j = 1, \dots , n$
	\begin{enumerate}
	    \item For $s = 1, \dots, N$
		First observation $z_{\mu}$ known: For each particle, at time 1, fill the invasion vector of $\vec{x}_j^t$ size N with $(x_{\mu})^1_j = 1$ and $(x_{s})^1_j = 0$ if $s \neq \mu$. 
		For each particle, at time 1, fill the invasion vector $\vec{z}_j^t$ of size N with $(z_{\mu})^1_j = 1$ and $(z_{s})^1_j = 0$ if $s \neq \mu$.
		\item Evaluate the importance weights up to a normalising constant:
		\[
		\tilde{w}^{1}_{j} = 1
		\]
	\end{enumerate}
	\item For $j = 1, \dots , n$ normalise the importance weights: 
	\[
	w^{1}_{j} = \frac{1}{n}
	\]
\end{enumerate}

 \State  \bf{Iterate:} \normalfont For $t$ from 2 to $T$

\begin{enumerate}
	\item For $j = 1, \dots , n$
	\begin{enumerate}
  		\item Sample $(r)^t_j \sim Bernoulli(\theta)$ and $(l)^t_j \sim Bernoulli(\theta)$. For $s = 1, \dots, N$ for each particle, at time t, fill the invasion vector $\vec{x}_j^t$ of size N with the values of $\vec{x}_j^{t-1}$ then substitute $(x_s)^t_j = (r)^t_j$ for $s$ s.t. $(x_{s-1})^{t-1}_j = 1$ and $(x_s)^{t-1}_j = 0$ and substitute $(x_s)^t_j = (l)^t_j$ for $s$ s.t. $(x_s)^{t-1}_j = 0$ and $(x_{s+1})^{t-1}_j = 1$.
  		\item Sample k elements $(a_k)^t_j \sim Bernoulli(\varphi)$ until the first 0 is sampled, and h elements $(b_h)^t_j \sim Bernoulli(\varphi)$ until the first 0 is sampled. For $s = 1, \dots, N$ for each particle, at time t, fill the observation vector $\vec{z}_j^t$ of size N with the values of $\vec{z}_j^{t-1}$ then substitute the k elements $(z_{s+k})^t_j = (r)^t_j$ for $s$ s.t. $(z_{s-1})^{t-1}_j = 1$ and $(z_s)^{t-1}_j = 0$ and substitute the h elements $(z_{s-h})^t_j = (l)^t_j$ for $s$ s.t. $(z_s)^{t-1}_j = 0$ and $(z_{s+1})^{t-1}_j = 1$.
		\item Correct every element of $\vec{x}^t_j$ in light of the data $\vec{z}^t$ to obtain new samples $(\vec{x'})^t_j$ directly substituting the new data: for $s = 1, \dots ,N$ if $(z^t)_s = 1$, $((x')_j^t)_s = 1$ else if $(z^t)_s = 0$, $((x')_j^t)_s = (x^t_j)_s$. Correct every element of $\vec{z}^t_j$ in light of the data $\vec{z}^t$ to obtain $(\vec{z'})^t_j$ simply substituting $(\vec{z'})^t_j$ with $\vec{z}^t$.
		\item Evaluate the weights up to a normalising constant using the values for $q$ from equation (\ref{eq:4}):
		\[
		\tilde{w}^{t}_{j} = \frac{q(\vec{(X')}^{t}_j, \vec{(z')}^{t}_j | \vec{Z}^{t-1})}{q(\vec{X}^{t}_j, \vec{z}^{t}_j | \vec{Z}^{t-1})}
		\]
	\end{enumerate}
	\item For $j = 1, \dots , n$ normalise the importance weights:
	\[
	w^{t}_{j} = \frac{\tilde{w}^t_j}{\sum_{k=1}^{n}\tilde{w}^{t}_k}
	\]
	\item Perform resampling
	\begin{enumerate}
	    \item Draw $n$ particles from the current particle set with probabilities proportional to their weights. Replace the current particle set with the new $n$ particles.
	    \item Set $w^t_j=1/n$.
	\end{enumerate}
\end{enumerate}
  
 \end{algorithmic}
\end{algorithm}

\section{proof of the identity in section \ref{sec:2} }

Define a probability measure $\P^*$ having density $p^*$ on $\Omega^*_t$ such that the marginal distribution of $\P^*$ on $\Omega_t'$ is $\P$, that is $\P = \P^* \circ \pi^{-1}$, where $\P$ is the measure with density $p$ on $\Omega_t'$. (Here the measurable sets in $\Omega_t$, $\Omega_t'$ and $\Omega_t^*$, are constructed from Borel sets on $\Real^M$ and subsets of $2^M$ in the manner implied by the sequence of cross-products and restrictions used to define $\Omega_t$, $\Omega_t'$ and $\Omega_t^*$. Similarly, reference measures are constructed from Lebesgue measure on $\Real^M$ and counting measure on $2^M$.)

By the disintegration theorem (see \cite{Rohlin}) there exists a family of measures $\{ \nu_{\vec{(y')}^t} \}_{\vec{(y')}^t \in \Omega'_t}$ on $\Omega^*_t$ such that for every measurable Borel function $m : \Omega^*_t \rightarrow [0, \infty]$
\begin{align*}
    &\int_{\Omega^*_t} m(\vec{(y')}^t, \vec{y}^t)\diff \P^* = \\
    & \int_{\Omega_t'} \bigg( \int_{\pi^{-1}(\vec{(y')}^t)} m(\vec{(y')}^t, \vec{y}^t) \diff \nu_{\vec{(y')}^t}  \bigg) \diff \P((\vec{y'})^t).
\end{align*}
It follows that for any event $A$,
\begin{align*}
    &\P(A) = \P^{*}(\pi^{-1}(A)) = \\
    &\int_{\Omega_t^*} I_A(\pi(\vec{(y')}^t, \vec{y}^t)) p^*(\vec{(y')}^t, \vec{y}^t | \vec{Z}^t) \D \P^*  = \\
    & \int_{\Omega_t'} I_A(\vec{(y')}^t) \Bigg[ \int_{\pi^{-1}(\vec{(y')}^t)} p^*(\vec{(y')}^t, \vec{y}^t | \vec{Z}^t) \D \nu_{\vec{(y')}^t} (\vec{y}^t) \Bigg] \D \P (\vec{(y')}^t)
\end{align*}
and hence
\begin{equation*}
    p(\vec{(y')}^t | \vec{Z}^t) = \int_{\pi^{-1}(\vec{(y')}^t)} p^*(\vec{(y')}^t, \vec{y}^t | \vec{Z}^t) \D \nu_{\vec{(y')}^t}(\vec{y}^t).
\end{equation*}
Moreover,
\begin{align*}
    &E_{p}[l(\vec{(y')}^{t})]  = \int_{\Omega_t'} l(\vec{(y')}^{t})\; p(\vec{(y')}^{t} | \vec{Z}^{t}) \D \P (\vec{(y')}^{t}) \\
    &= \int_{\Omega_t'} l(\vec{(y')}^{t})\Bigg[\int_{\pi^{-1}(\vec(y'))} p^*(\vec{(y')}^{t}, \vec{y}^{t} | \vec{Z}^{t}) \D \nu_{\vec{(y')}^t}(\vec{y}^t) \Bigg] \D \P \vec{(y')}^{t} \\ 
    &= \int_{\Omega_t^*} l(\pi(\vec{(y')}^{t}, \vec{y}^{t}))\; p^*(\vec{(y')}^{t}, \vec{y}^{t} | \vec{Z}^{t}) \D \P^*(\vec{(y')}^t, \vec{y}^t) \\ 
    &= E_{p^*}[l(\pi (\vec{(y')}^{t}, \vec{y}^{t}))].
\end{align*}

\end{document}